\magnification=\magstep1
\baselineskip=24 true pt
\hoffset=0.0 true cm
\voffset=0.5 true cm
\hsize=17.0 true cm
\vsize=21.0 true cm
\parindent=30 true pt
\parskip=9 true pt
\raggedbottom

\centerline{\bf THE LEVI-CIVITA SPACETIME}
\bigskip
\bigskip
\bigskip
{\baselineskip =14 true pt \centerline{\bf M. F. A. da Silva}
\centerline{\tenit Departamento de F\'{\i}sica e Qu\'{\i}mica,
Universidade Estadual Paulista,}
\centerline{\tenit Av. Ariberto Pereira da Cunha 333, 12500,
Guaratinguet\' a, SP, Brazil and}
\centerline{\tenit Departamento de Astrof\'{\i}sica,
CNPq-Observat\'orio Nacional,}
\centerline{\tenit Rua General Jos\' e Cristino 77, 20921-400 Rio de
Janeiro - RJ, Brazil.}}
\bigskip
{\baselineskip =14 true pt \centerline{\bf L. Herrera}
\centerline{\tenit Departamento de F\'{\i}sica, Facultad de Ciencias,} 
\centerline{\tenit Universidade Central de Venezuela and} 
\centerline{\tenit Centro de F\'{\i}sica, Instituto Venezolano de 
Investigaciones Cient\'{\i}ficas,}
\centerline{\tenit Caracas, Venezuela. Postal address: Apartado 80793,
Caracas 1080A, Venezuela}}
\bigskip
{\baselineskip =14 true pt \centerline{\bf F. M. Paiva and N. O.
Santos} \centerline{\tenit Departamento de Astrof\'{\i}sica,
CNPq-Observat\'orio Nacional,}
\centerline{\tenit Rua General Jos\'e Cristino 77, 20921-400 Rio de
Janeiro - RJ, Brazil}} 

\bigskip
\bigskip
\noindent Author's internet addresses respectively: mfas@on.br, 
lherrera@conicit.ve, fmpaiva@on.br and nos@on.br. 

\vfill\eject

{\bf Abstract}
\noindent

We consider two exact solutions of Einstein's field equations
corresponding to a cylinder of dust with net zero angular momentum. In
one of the cases, the dust distribution is homogeneous, whereas in the
other, the angular velocity of dust particles is constant [1]. For both
solutions we studied the junction conditions to the exterior static
vacuum Levi-Civita spacetime. From this study we find an upper limit
for the energy density per unit length $\sigma$ of the source equal
${1\over 2}$ for the first case and ${1\over 4}$ for the second one.
Thus the homogeneous cluster provides another example [2] where the
range of $\sigma$ is extended beyond the limit value ${1\over 4}$
previously found in the literature [3,4]. Using the Cartan Scalars
technics we show that the Levi-Civita spacetime gets an extra symmetry
for $\sigma={1\over 2}$ or ${1\over 4}$. We also find that the cluster
of homogeneous dust has a superior limit for its radius, depending on
the constant volumetric energy density $\rho_0$.
\bigskip
\bigskip
\bigskip

{\bf 1. Introduction}
\noindent

The Levi-Civita metric [5] is the most general cylindrical static
vacuum metric. It will be used as the exterior spacetime of static
cylindrical sources.

Some non-vacuum exact solutions with cylindrical symmetry may be found
in the literature. One very simple solution is the cluster of
particles. This source is constituted by a great number of small
gravitational particles which move freely under the influence of the
field produced by all of them together. The first model of cluster was
presented by Einstein [6] in spherical symmetry. Raychaudhuri and Som
[7], based on the Einstein's ideas, obtained the first cylindrical
analog case. They considered the source filled with dust with an equal
number of particles moving in clockwise and anticlockwise directions
and found some special cylindrical solutions. They also deduced a
relation between the gravitational mass and the sum of the free
particles masses of their clusters.  Another cluster solution was
obtained by Teixeira and Som [1]. They suposed a constant angular
velocity for the dust particles. From the Teixeira and Som solution,
Lathrop and Orsene [4] found an upper limit for the linear mass density
of th

In 1969 Gautreau and Hoffman [9] demonstrated that there is no timelike
circular geodesic in the Levi-Civita metric if $\sigma \geq {1\over
4}$. Based on this, Bonnor and Martins [3] conjectured that the
Levi-Civita metric does not represent an infinite line mass if
$\sigma\geq{1\over 4}$. Later, Bonnor and Davidson [2] presented a
cylindrical source, filled with perfect fluid, and showed that the
matching of this source with the Levi-Civita metric permits values of
$\sigma>{1\over 4}$, but $<{1\over 2}$.  In another work, Stela and
Kramer [10] found a source with $\sigma\buildrel < \over\sim 0.35$
using a numerical interior solution.

Inspired by the Teixeira and Som solution, and imposing the constance
of the energy density of the source, instead of the constance of the
angular velocity, we found an exact solution for an homogeneous
cylindrical cluster.  For both these clusters we studied the junction
conditions to the exterior static vacuum Levi-Civita spacetime. We
found a superior limit for the linear energy density $\sigma$ of the
source equal ${1\over 4}$ for the first case and ${1\over 2}$ for the
second one. The limit obtained for the Teixeira and Som cluster is in
accordance with the limit found by Lathrop and Orsene [4]. The range of
$\sigma$ for the homogeneous cluster solution, extends the range found
by Bonnor and Davidson [2], for a perfect fluid. Our solution allows
$\sigma={1\over 2}$.

The paper is organized as follows. In the next section we present the
static cylindrical spacetime in general. In the section three we give
the notation and the spacetime at the exterior of the boundary of the
source. In the fourth section we describe the interior spacetime. This
section is subdivided into two subsections, where we present a new
cylindrical cluster solution for homogeneous dust, found by us, and the
Teixeira and Som [1] cluster solution. The junction conditions of these
clusters to the Levi-Civita spacetime are also included in this
section. In the conclusion we sum up our main results. In the appendix
we analyse the symmetries of the Levi-Civita spacetime using the Cartan
scalars approach.
\bigskip
\bigskip
\bigskip

{\bf 2. Spacetime}

\noindent
The spacetime is described by the general cilindrically symmetric
static metric $$ds^2=-fdt^2+e^\mu(dr^2+dz^2)+ld\varphi^2,\eqno (2.1)$$
where $f$, $\mu$ and $l$ are functions only of $r$. The ranges of the
coordinates $t$, $z$ and $\varphi$ are $$-\infty<t<\infty, \quad
-\infty<z<\infty, \quad 0\leq\varphi\leq 2\pi,\eqno (2.2)$$ and the
hypersurface $\varphi=0$ and $\varphi=2\pi$ being identified. The
coordinates are numbered $$x^0=t, \quad x^1=r, \quad x^2=z, \quad
x^3=\varphi.\eqno (2.3)$$ We shall impose the Einstein's field
equations $$R_{\mu\nu}=k\left(T_{\mu\nu}-{1\over
2}g_{\mu\nu}T\right).\eqno (2.4)$$

The non zero components of $R_{\mu\nu}$ for the metric (2.1) are $$2
e^\mu D R^0_0=\left({lf^\prime\over D} \right)^\prime,\eqno(2.5)$$  $$2
e^\mu D R^3_3= \left({fl^\prime\over D} \right)^\prime,\eqno(2.6)$$ $$2
R_{11}=-\mu^{\prime\prime} +\mu^\prime {D^\prime \over D} -
2{D^{\prime\prime} \over D} + {f^\prime l^\prime \over
D^2},\eqno(2.7)$$ $$2R_{22}=-\mu^{\prime\prime} -\mu^\prime{D^\prime
\over D},\eqno(2.8)$$ where the primes stand for differentiation with
respect to $r$ and $$D^2=fl.\eqno (2.9)$$

For the line element (2.1) we found that the circular geodesics
equations are $$\dot r=\dot z =0,\eqno (2.10) $$
$$l^\prime\dot\varphi^2-f^\prime\dot t^2 =0, \eqno (2.11)$$ $$\ddot t
=0,\eqno  (2.12)$$ $$\ddot\varphi =0,\eqno (2.13)$$ $$\left({ds\over
dt}\right)^2 =l\omega^2-f. \eqno (2.14)$$ where the dot stands for
differentiation with respect to $s$, and the angular velocity of the
particle $\omega=\left(dx^3/ ds\right)\left(dx^0/ds\right)^{-1}$, is
$$\omega^2=\left({d\varphi\over dt}\right)^2={f^\prime\over
l^\prime}.\eqno (2.15)$$ For a stationary spacetime the normal velocity
of the particle defined as the change in the displacement normal to
$\tau^\mu=\{1,0,0,0\}$ relative to its displacement parallel to
$\tau^\mu$, where $\tau^\mu$ is a timelike Killing vector, is [11]
$$W^\mu \buildrel\rm def\over
=\left[\sqrt{-g_{00}}\left(dx^0+{g_{0a}\over
g_{00}}dx^a\right)\right]^{-1}V^\mu,\eqno(2.16)$$ where
$$V^\mu\buildrel\rm def\over = \left(-{g_{0a}\over
g_{00}}dx^a,dx^1,dx^2,dx^3\right)$$ Latin indexes range from 1 to 3.
So, for the static metric (2.1), where $g_{0a}=0$, the three velocity
associated to the particle, defined by $W^2=W^\mu W_\mu$, is
$$W^2={l\over f}\omega^2={lf^\prime\over fl^\prime}.\eqno (2.17)$$

Considering equations (2.15) and (2.17) the geodesic equation (2.14)
can be written as $$\left({ds\over dt}\right)^2=(W^2-1)f.\eqno (2.18)$$
The above equation shows that circular geodesics are timelike, null or
spacelike for respectively $W<1$, $W=1$ and $W>1$.

The spacetime is divided into two regions: the interior, with $0\leq r
\leq R$, to a cylindrical surface of radius $R$ centered along $z$; and
the exterior, with $R\leq r<\infty$.  On the boundary surface $r=R$ the
first and second fundamental forms have to be continuous [12]. Choosing
the same coordinates for the exterior and interior spacetimes these
conditions become $$\left[g^-_{\mu\nu}-g^+_{\mu\nu}
\right]_\Sigma=0,\eqno(2.19)$$ $$\left[g^{- \prime}_{\mu\nu}-g^{+
\prime}_{\mu\nu} \right]_\Sigma=0,\eqno(2.20)$$ where the indexes $-$
and $+$ stand for the interior and exterior spacetimes respectively.  
\bigskip
\bigskip
\bigskip

{\bf 3. Exterior spacetime}
\noindent

In this section we present the exterior spacetime and discuss the
symetries.

The exterior spacetime is filled with vacuum, hence Einstein's
equations (2.4) reduce to $R_{\mu\nu}=0$.  The general solution for
(2.5)-(2.8) is the static Levi-Civita metric, $$f=ar^{4\sigma},\quad
e^\mu={1\over a} r^{4\sigma (2\sigma-1)}, \quad l={1\over a}
r^{2(1-2\sigma)}, \eqno (3.1)$$ where $a$ and $\sigma$ are constants.
The parameter $a$ is associated with the angular deffect [13] while the
parameter $\sigma$ can be interpreted as the linear energy density of
the source [13,14].

We shall now study some properties of the circular geodesics in this
spacetime.  A test particle in circular geodesics has angular velocity
(2.15) and three velocity (2.17) given by $$\omega^2={2\sigma\over
1-2\sigma}ar^{2(4\sigma-1)},\eqno (3.2)$$ $$W^2={2\sigma\over
1-2\sigma}.\eqno (3.3)$$
From equation (2.14) we see that
$$\left({ds\over dt}\right)^2=-{1-4\sigma\over
1-2\sigma}ar^{4\sigma}.\eqno (3.4)$$ Thus, circular geodesics are
timelike, null or spacelike for respectively $0\leq\sigma<{1\over 4}$,
$\sigma={1\over 4}$ and ${1\over 4}<\sigma\leq{1\over 2}$. The limit
$\sigma={1\over 2}$ implies $W\rightarrow\infty$.
\bigskip
\bigskip
\bigskip

{\bf 4. Interior spacetime}

\noindent
The interior spacetime is described by a cylinder filled with a
rotationally symmetric cluster of dust with zero net angular momentum.
The energy momentum tensor is $$T^\mu_\nu={1\over 2}\rho(u^\mu
u_\nu+v^\mu v_\nu),\eqno (4.1)$$ where $\rho$ is the energy density and
$u^\mu$ and $v^\mu$ are the four velocities
$$u^\mu=(u^0,0,0,\omega),\quad v^\mu=(u^0,0,0,-\omega), \eqno (4.2)$$
satisfying $$u^\mu u_\mu=v^\mu v_\mu=-1.\eqno (4.3)$$
From Einstein's equations (2.4) and from (2.5)-(2.8) and (4.1) we
obtain
$$fl=r^2,\eqno(4.4)$$ $$\mu^\prime=-{f^\prime\over
f}\left(1-{rf^\prime\over 2f}\right), \eqno (4.5)$$ $$k\rho r
e^\mu=-(r\mu^\prime)^\prime,\eqno (4.6)$$ $$W^2=\left({\omega r\over
f}\right)^2={rf^\prime\over 2f}{1\over \left(1-{rf^\prime\over
2f}\right)},\eqno (4.7)$$ where we have computed in (4.7) the three
velocity (2.17) of a particle in the cluster. Note that, although the
particles of the cluster are rotating, this source, with null net
angular momentum, generates a static  spacetime.

\bigskip
{\bf 4.1. Cluster of homogeneous dust}

\noindent
Considering a homogeneous distribution of dust inside the cylinder
$0\leq r\leq R$ we have $$\rho=\rho_0={\rm constant}.\eqno (4.1.1)$$
The solution of (4.5)-(4.6) with (4.1.1) suitable for a matching to the
Levi-Civita metric is
$$\eqalignno{f={1\over\sqrt{2}}&\left(1-3br^2+
\sqrt{(1+br^2)(1-7br^2)}\right)^{1\over
2}\times\cr &\exp\left[{7\over
2\sqrt{7}}\left(\arcsin\left[-{(3+7br^2)\over
4}\right]-\arcsin{\left[-{3\over
4}\right]}\right)\right],&(4.1.2)\cr}$$ $$e^\mu=(1+br^2)^{-2},\eqno
(4.1.3)$$ where we imposed that the geometry is euclidian on the
rotation axis and $b\buildrel\rm def\over =k\rho_0/8$.
From (4.7) and (4.1.2) we have for the three velocity of the dust
particles
$$W^2={\sqrt{1+br^2}-\sqrt{1-7br^2}\over
\sqrt{1+br^2}+\sqrt{1-7br^2}}.\eqno (4.1.4)$$
From equation (4.1.2) we can see that there is a restriction on the
radius of the orbit of the homogeneous cluster particles, for a given
gravitational mass per unit length of the cylinder. More specificaly,
we should have that  $$r^2\leq {8\over 7k\rho_0},\eqno (4.1.5)$$
in order that the metric potential $f$ could be real. Equation (4.1.4)
shows that in this limit the three velocity of the dust particles is
equal to the unit. A similar behaviour is also found in the van Stockum
solution [15,16].

Considering the matching between the interior and exterior spacetimes,
given by (2.19) and (2.20), we obtain $$\sigma={1\over
4}\left(1\pm\sqrt{{1-7bR^2\over 1+bR^2}}\right),\eqno (4.1.6)$$
$$a={b^2R^{4(2\sigma^2-\sigma+1)}
\over\sigma^2(2\sigma-1)^2}=(1+bR^2)^2R^{-{4bR^2\over 1+bR^2}}.\eqno
(4.1.7)$$

Equation (4.1.6) imposes a superior limit on the radius of the source's
boundary, independently of the choice of the sign before the square
root. This limit is identical to equation (4.1.5).  Furthermore,
(4.1.5) and (4.1.6) implies that there are two possible ranges for
$\sigma$, depending on the choice of the sign before the square root.
If we choose the negative sign we have $0\leq \sigma \leq {1\over 4}$,
with $\rho_0=0$ corresponding to $\sigma=0$, while if we choose the
positive sign, then ${1\over 4}\leq\sigma\leq{1\over 2}$, with
$\rho_0=0$ corresponding to $\sigma={1\over 2}$. The value
$\sigma={1\over 4}$ is included in both solutions and corresponds to
$W=1$ from equation (4.1.4). Observe that the limits imposed by
equations (3.3) and (3.4) are related with the existance of circular
geodesics for test particles in the Levi-Civita spacetime, while the
above limits arise from the junction conditions. Although the junction
conditions (4.1.6) and (4.1.7) do not impose any restriction on the
value $\sigma={1\over 4}$, equation (4.1.4), as we had seen, shows that
when $\sigma={1\over 4}$, which means $R^2=8/(7k\rho_0)$, the three
velocity of the particles is equal to the unit. So, the partic

\bigskip
{\bf 4.2. Cluster of constant rotating dust}

\noindent
We consider now the Teixeira and Som solution where
$$\omega=\omega_0={\rm constant},\eqno (4.2.1)$$
given by
$$f={1\over 2}\left[1+(1+4\omega_0^2r^2)^{1\over 2}\right],\eqno (4.2.2)$$
$$e^\mu=(1+4\omega_0^2r^2)^{-{1\over 4}},\eqno (4.2.3)$$
$$2\pi\rho={\omega_0^2\over(1+4\omega_0^2r^2)^{7\over 4}},\eqno (4.2.4)$$
$$W^2={4\omega_0^2r^2\over (1+\sqrt{1+4\omega_0^2 r^2})^2}.\eqno (4.2.5)$$
Now considering the matching (2.19)-(2.20) we obtain
$$\sigma={1\over 4}\left(1-{1\over 
\sqrt{1+4\omega_0^2 R^2}}\right),\eqno (4.2.6)$$
$$a=R^{4\sigma(2\sigma-1)}(1+4\omega_0^2R^2)^{1\over 4}.\eqno (4.2.7)$$

We shall now state some properties of this matching.  We can see that
if $\sigma=0$ equations (4.2.6)-(4.2.7) give $R\omega_0=0$ and $a=1$
producing the Minkowski spacetime as expected. Note that the solution
is well behaved for $R=0$ since this implies $\sigma =0$ (i.e., finite
density energy per unit length of the infinite line mass). On the other
hand, even when $R=0$, from equation (4.2.4), the density energy $\rho$
of the source vanishes only if we have $\omega_0=0$.

Equation (4.2.6) shows that $$\lim_{R\rightarrow\infty}\sigma={1\over
4},\eqno(4.2.8)$$ and $$\lim_{R\rightarrow\infty}w_0=1.\eqno (4.2.9)$$
So, ${1\over 4}$ represents a superior limit for $\sigma$ in order that
the interior solution generates an exterior Levi-Civita spacetime.
Lathrop and Orsene [4] found the same limit for the linear mass density
of the Teixeira and Som cluster solution. They used a definition for
this parameter given by Vishveshwara and Winicour [8]. This source,
nevertheless, do not present a limitation on the radius of its
boundary. As in the case of the homogeneous cluster, here the value
${1\over 4}$ for the parameter $\sigma$ implies that the particles of
the cluster are travelling with the speed of light, as can be seen from
the equation (4.2.5). So, we again need to avoid the value
$\sigma={1\over 4}$ or otherwise consider the possibility of
counter-rotating photons.

\bigskip
\bigskip
\bigskip

{\bf 5. Conclusion}

\noindent
We found the exact solutions of Einstein's field equations for an
homogeneous cylinder constituted by an equal number of particles of
dust moving in clockwise and anticlockwise directions. The matching of
this source with the static vacuum of Levi-Civita is allowed only for a
specific range of the linear energy density parameter, that is
$0\leq\sigma\leq{1\over 2}$. This extends the range of $\sigma$ found
by Bonnor and Davidson [2] for a perfect fluid source. We also found
that, for a given volumetric energy density, this source presents an
upper limit for its radius. In the literature there is, at least, the
van Stockum solution as another cylindrical example in which there is a
limit for the radius of the source depending on its volumetric density
[15,16]. In this case, the limitation on the radius comes out in order
to avoid a change in the signature of the metric.

Using the solution for a cluster constituted by dust particles with
constant angular velocity and zero net angular momentum, obtained
firstly by Teixeira and Som [1], and matching it with the Levi-Civita
metric, we found that the parameter $\sigma$ in this case should be
smaller or equal than ${1\over 4}$. Considering $\sigma$ as the
gravitational mass per unit length this limit is in agreement with the
result of Lathrop and Orsene [4]. While for the Teixeira and Som
solution the matching does not present a limitation to the radius of
the source, for our homogeneous cluster solution the matching imposes a
superior limit for its radius, depending on the volumetric energy
density $\rho_0$.

We note from equation (3.4) that circular geodesics in the Levi-Civita
spacetime become null when $\sigma$ is equal ${1\over 4}$. Some authors
[Gautreau \& Hoffmann [3,9] use this fact as a restriction for the
linear density of the source. Nevertheless, as pointed out by [7], this
result is similar to the Newtonian cylindrical analog case, i. e., for
a higher density cylinder, all particles (with speed less than the
light) should fall. However this argument is not without problems,
since, as showed by Bonnor and Martins [3], in the interval ${1\over
4}\leq\sigma\leq{1\over 2}$ the gravitational field seems to get weaker
as $\sigma$ increases [17]. Note that $\sigma={1\over 2}$ means that
there is no any matter inside the cylinder ($\rho_0=0$ and the
spacetime is locally flat, in accordance to the Cartan scalars). From
equation (4.1.6) we can see that when we choose the positive sign, the
parameter $\sigma$ grows up while the volumetric density $\rho_0$
decreases.

The analysis of the Cartan scalars, in the appendix, summarizes the
symmetry properties of the Levi-Civita metric. It is in general a
Petrov type I metric and becomes a Petrov type D metric when $\sigma
=-{1\over 2}, 0, {1\over 4}, {1\over 2}$ and $1$. For $\sigma=0$ or
${1\over 2}$ the metric becomes flat, which is in accordance with our
cluster solution since when $\sigma=0$ or ${1\over 2}$ the volumetric
energy density $\rho_0$ of the cluster vanishes. For $\sigma=-{1\over
2}, {1\over 4}, 1$ the metric gets one extra symmetry.

\bigskip
\bigskip
\bigskip

{\bf Appendix}
\noindent

It is known [18] that the so called 14 algebraic
invariants (and even all the polinomial invariants of any order) are
not sufficient for locally characterizing a spacetime, in the sense
that two metrics may have the same set of invariants and be not
equivalent. As an example, all these invariants vanish for both
Minkowski and plane-wave [18,19]
spacetimes and they are not the same. A complete local characterization
of spacetimes may be done by the Cartan scalars. Briefly, the Cartan
scalars are the components of the Riemann tensor and its covariant
derivatives, up to possibly the $10^{th}$ order, calculated in a
constant frame. Nevertheless, for a complete characterization of a
spacetime, up to the $3^{th}$ order is usually sufficient. For a
review, see [13] and references
therein.  In practice, the Cartan scalars are calculated using the
spinorial formalism. For the purpose here, the relevant quantities are
the Weyl spinor $\Psi_A,$ and its first and second covariant
symmetrized derivatives $\nabla\Psi_{AW'}$ and $\nabla^2\Psi_{AW'}$,
which represent the Weyl tensor and its covariant derivatives. It
should be stressed that, although the Cartan scalars provide a local
characterization of the spacetime, global properties such as
topological deffects do not probably appear in them.

The Cartan scalars for the Levi-Civita metric coincide with those for
the Weyl class of the Lewis metric and were given in
[13]. Using the null frame 
$$\eqalign{
&\omega^0 = {1\over\sqrt{2}}(\theta^0 +   \theta^1),\qquad \omega^1 
= {1\over\sqrt{2}}(\theta^0 -   \theta^1)\cr 
 &\omega^2 = {1\over\sqrt{2}}(\theta^2 + i \theta^3),\qquad \omega^3 
= {1\over\sqrt{2}}(\theta^2 - i \theta^3).\cr}\eqno (A.1)$$ 
where $\theta^A$ is an orthonormal frame given by:
$$\theta^0 = r^{2\sigma} dt,\quad  
\theta^1 = {1\over a}r^{(1-2\sigma)} d\varphi,  \quad
\theta^2 = r^{(4\sigma^2-2\sigma)} dr,  \quad
\theta^3 = r^{(4\sigma^2-2\sigma)} dz,\eqno (A.2)$$
the non-vanishing Cartan scalars become
$$\eqalign{\Psi_2 & = -(2\sigma-1)\sigma r^{4\sigma-8\sigma^2-2}\cr
\Psi_4 = \Psi_0 & = (4\sigma-1)\Psi_2 \cr
\nabla\Psi_{01'} =\nabla\Psi_{50'} & =\sqrt{2}(8\sigma^2-4\sigma+1)
(4\sigma-1)(2\sigma-1)\sigma r^{6\sigma-12\sigma^2-3}\cr
\nabla\Psi_{10'} = \nabla\Psi_{41'} & =
\sqrt{2}(4\sigma-1)(2\sigma-1)\sigma  r^{6\sigma-12\sigma^2-3}\cr
\nabla\Psi_{21'} = \nabla\Psi_{30'} & = 
\sqrt{2}(4\sigma^2-2\sigma+1)(2\sigma-1)
\sigma r^{6\sigma-12\sigma^2-3}.}\eqno (A.3)$$ 
In order to discuss the symmetries, we list some quantities which are
important in the Petrov classification [20]
$$D = 64(4\sigma-1)^2(2\sigma+1)^2(\sigma-1)^2\Psi_2^6,\eqno (A.4)$$ 
$$I = ((4\sigma-1)^2 + 3)\Psi_2^2, \eqno (A.5)$$
$$J = ((4\sigma-1)^2 - 1)\Psi_2^3, \eqno (A.6)$$
$$G = 0, \eqno (A.7)$$
$$N \buildrel\rm def \over = \Psi_0I-12H^2 = 
(4\sigma-1)^2((4\sigma-1)^2-9)\Psi_2^4.\eqno (A.8)$$
The metric is Petrov type I unless $D = 0$. This can happen if and
only if $\sigma = -1/2$, $0$, $1/4$, $1/2$ or $1$. For $\sigma = 0$ or
$1/2$, we have from (A.3) that $\Psi_2 = 0$ and therefore the
metric is flat. For $\sigma = 1/4$, we have from (A.3) that
$\Psi_0 = \Psi_4 = 0$ and therefore the metric is Petrov type D. For
$\sigma = -1/2$ or $1$, we see that $I$ and $J$ are different from zero
and that $G = N = 0$, which also characterize Petrov type D metrics.

As it is well known, the Minkowski spacetime has the Poincar\' e group
as its isometry group and the Lorentz group as its isotropy subgroup.
For a general spacetime, the isometry and isotropy groups are subgroups
of the Poincar\' e and Lorentz groups respectively. The Cartan scalars
(A.3) do not depend on $t$, $\varphi$ and $z$, showing that the
isometry group is at least 3 dimensional. Concerning the isotropies, it
is known [21] that Petrov type I, II or III metrics have none, Petrov
type D and N may have up to 2 and Petrov type 0 metrics may have all 6
isotropies.  Therefore, the Levi-Civita metric has in general no
isotropies.  Isotropies may arise for $\sigma = -1/2$, $0$, $1/4$,
$1/2$ or $1$. In fact, for $\sigma = 0$ or $1/2$, there are 6
isotropies, since the metric becomes flat and for $\sigma = -1/2$,
$1/4$ or $1$, i.e., the Petrov type D cases, we shall show now that the
metric gets an extra isotropy and therefore the spacetime has more
symmetries than the original cylindrical symmetry.

The case $\sigma = 1/4$ can be easily analysed, since the 
Cartan scalars
will be still in a canonical basis [22]. 
In fact, from
(A.3) we get that the non-zero Cartan scalars become 
$$\eqalign{\Psi_2 &= 1/8r^{-3/2},\cr
\nabla\Psi_{21'} = \nabla\Psi_{30'} &= -3\sqrt{2}/(32)r^{-9/4}, \cr
\nabla^2\Psi_{22'} = \nabla^2\Psi_{40'} = 3/(16)r^{-3}&,
\quad \nabla^2\Psi_{31'} = 27/(128)r^{-3}.}\eqno (A.9)$$ 
Actually, $\nabla^2\Psi_{AW'}$ does not come from (A.3) since in
that case the second derivative was not necessary. Therefore, the
second derivative was, here, calculated directly from the tetrad frame
(using SHEEP and CLASSI [23,24]).  The
isotropy group corresponding to this set of Cartan scalars is a
one-parameter group of boosts in the $\omega^0-\omega^1$ plane which,
from (A.1) and (A.2), involves the $t$ and $\varphi$ coordinate axis.

The two other cases are more complicate since the Cartan scalars will
not be in a canonical basis for Petrov type D. 
The new canonical basis and the Cartan scalars in these cases are, 
for $\sigma = 1$, 
$$\eqalign{\omega^0 ={\sqrt{2}\over 4}r^2dt + 
{\sqrt{2}\over 4}r^{2}dz &,\quad
\omega^1 = \sqrt{2}r^2 dt - \sqrt{2}r^ 2dz, \cr
\omega^2 = 
{1\over\sqrt{2}}ia^{-1/2}r^{-1}d\varphi - 
{1\over\sqrt{2}}r^2dr &,\quad
\omega^3 = 
-{1\over\sqrt{2}}ia^{-1/2}r^{-1}d\varphi - 
{1\over\sqrt{2}}r^2dr, \cr
\Psi_2 &= 2r^{-6}, \cr
\nabla\Psi_{21'} = \nabla\Psi_{30'} &= 6\sqrt{2}\,r^{-9}, \cr
\nabla^2\Psi_{22'} = \nabla^2\Psi_{40'} &=  48r^{-12} ,\quad
\nabla^2\Psi_{31'} = 54r^{-12},}\eqno (A.10)$$
and, for $\sigma = -1/2$, 
$$\eqalign{\omega^0 = {1\over\sqrt{2}}r^{-1}dt + 
{1\over\sqrt{2}}r^2dr  &,\quad
\omega^1 =  {1\over\sqrt{2}}r^{-1}dt -  {1\over\sqrt{2}}r^2dr, \cr
\omega^2 =  
-{1\over\sqrt{2}}a^{-1/2}r^2d\varphi + {1\over\sqrt{2}}ir^2dz &,\quad
\omega^3 =  
-{1\over\sqrt{2}}a^{-1/2}r^2d\varphi -  {1\over\sqrt{2}}ir^2dz, \cr
\Psi_2 &= 2r^{-6}, \cr
\nabla\Psi_{20'} = - \nabla\Psi_{31'} &=  -6\sqrt{2}\,r^{-9}, \cr
\nabla^2\Psi_{20'} = \nabla^2\Psi_{42'} &= 48r^{-12} ,\quad
\nabla^2\Psi_{31'} = -54r^{-12}.}\eqno (A.11)$$ 
The isotropy group corresponding to the set of Cartan scalars for
$\sigma = 1$ is the one-parameter group of boosts in the
$\omega^0-\omega^1$ plane which, from (A.10), involves the $t$ and $z$
coordinate axis and for $\sigma = -1/2$ it is the one-parameter group
of spatial rotations in the $\omega^2-\omega^3$ plane which, from
(A.11), involves the $\varphi$ and $z$ coordinate axis (this last case
is also discussed by [25]).

The non zero algebraic invariants of the Riemann tensor for (3.1) are
$$R_{\alpha\beta\gamma\delta}R^{\alpha\beta\gamma\delta}=64
\sigma^2(2\sigma-1)^2(4\sigma^2-2\sigma+1)
r^{-4(4\sigma^2-2\sigma+1)},\eqno
(A.12)$$
$$R_{\alpha\beta\gamma\delta}R^{\gamma\delta\mu\nu}R_{\mu\nu}^{\alpha\beta}
=768\sigma^4(2\sigma-1)^4
r^{-6(4\sigma^2-2\sigma+1)}.\eqno(A.13)$$ According to the Cartan
scalars and to (A.12) and (A.13) the metric (3.1) has infinite
curvature only at $r=0$ for all $\sigma$ except $\sigma=0$ and ${1\over
2}$ where the spacetime is flat.

\bigskip
\bigskip
\bigskip

{\bf Acknowledgment}
\noindent

MFAS and FMP gratefully acknowledge financial assistence from CAPES and
CNPq, respectively.

\bigskip
\bigskip
\bigskip

{\bf References}

\bigskip
\noindent
[1]Teixeira, A. F. F. \& Som, M. M. (1974), Il Nuovo Cimento {\bf 21 B}, 64.
\bigskip
\noindent
[2]Bonnor, W. B. \& Davidson, W. (1992), {\it Class. Quant. Grav.} 
{\bf 9}, 2065.
\bigskip
\noindent
[3]Bonnor, W. B. \& Martins, M. A. P. (1991), 
{\it Class. Quant. Grav.}
{\bf 8}, 727.
\bigskip
\noindent
[4]Lathrop, J. D. \& Orsene, M. S. (1980), J. Math. Phys. {\bf 27}(1), 152.
\bigskip
\noindent
[5]Levi-Civita, T. (1917), Rend. Acc. Lincei {\bf 26}, 307.
\bigskip
\noindent
[6]Einstein, A. (1939) Ann. of Math. {\bf 40}, 921.
\bigskip
\noindent
[7]Raychaudhuri, A. K. \& Som, M. M. (1962), Proc. Camb. Phyl. Soc. 
{\bf 58},
 338.
\bigskip
\noindent
[8]Vishveshwara, C. V. \& Winicour, J. (1977), J. Math. Phys. {\bf 18}, 1280.
\bigskip
\noindent
[9]Gautreau, R. \& Hoffman, R. B. (1969), Nuovo Cimento, {\bf B61}, 411.
\bigskip
\noindent
[10]Stela, J. \& Kramer, D. (1990), Acta Phys. Pol. {\bf B21}, 843.
\bigskip
\noindent
[11]Anderson, J. L., (1967), {\it Principles of Relativity, sect. 10.6(a)}
(New York, NY).
\bigskip
\noindent
[12]Darmois, G. (1927), {\it M\' e\-mo\-rial des Sciences 
Ma\-the\-ma\-ti\-ques} (Gauthier\- - \-Villars, Paris), 
Fasc. 25.
\bigskip
\noindent
[13]da Silva, M. F. A., Herrera, L.,
Paiva, F. M. \& Santos, N. O. (1994), prepint.
\bigskip
\noindent
[14]Marder, L. (1958), Proc. Roy. Soc. London, Ser A, {\bf 244}, 524.
\bigskip
\noindent
[15]van Stockum, W. J. (1937), Proc. R. Soc. Edin. {\bf 57}, 135.
\bigskip
\noindent
[16]Tipler, F. J. (1974), Phys. Rev. D, {\bf 9}, 2203.
\bigskip
\noindent
[17]Bonnor, W. B., private communication.
\bigskip
\noindent
[18]MacCallum, M. A. H. \& Skea, J. E. F.
(1992), SHEEP:  {\it A Computer Algebra System for General
Relativity}, in Algebraic Computing in General Relativity: Lecture
Notes from the First Brazilian School on Computer Algebra, Vol. 2,
edited by M. J. Rebou\c{c}as and W. L.  Roque, Oxford U. P., Oxford.
\bigskip
\noindent
[19]Schmidt, H-J (1994), prepint.
\bigskip
\noindent
[20]D'Inverno, R. A. \& Russell-Clark,
R. A. (1971), J.
Math. Phys. {\bf 12}, 1258--1263.
\bigskip
\noindent
[21]Karlhede, A. (1980), Gen. Rel.
Grav.  {\bf 12}, 693--707.
\bigskip
\noindent
[22]Paiva, F. M. (1993), {\it Limites de
espa\c{c}os-tempos em gravita\c{c}\~ao} Tese de Doutorado, Centro
Brasileiro de Pesquisas F\'{\i}sicas, Rio de Janeiro.
\bigskip
\noindent
[23]\AA man, J. E. (1987), {\it Manual for CLASSI -
Classification Programs for Geometries in General Relativity} (Third
Provisional Edition), University of Stockholm Report.
\bigskip
\noindent
[24]Frick, I. (1977), {\it SHEEP Users Guide},
Institute of Theoretical Physics, University of Stockholm Report
77--14. 
\bigskip
\noindent
[25]Bedran, M. L., Calv\~ao, M. O.,
Paiva, F. M. \& Soares, I. D. (1994), prepint.
\bigskip
\noindent

\end